\documentclass{iaus}
\usepackage{graphicx}

\title[The Shining Future of UV Spectral Synthesis] 
{The Shining Future of UV Spectral Synthesis}

\author[Pellerin \& Finkelstein]   
{Anne Pellerin \and Steven L. Finkelstein}

\affiliation{George P. and Cynthia W. Mitchell Institute for Fundamental Physics and Astronomy, Department of Physics, Texas A\&M University, College Station, TX 77843, USA \\ emails: {\tt pellerin@physics.tamu.edu, stevenf@physics.tamu.edu}}

\pubyear{2010}
\volume{262}  
\pagerange{TBD}
\jname{Stellar Populations: Planning for the Next Decade}
\editors{G. Bruzual \& S. Charlot, eds.}

\begin{document}

\maketitle

\begin{abstract}
With the coming generation of instruments and telescopes 
capable of spectroscopy of high redshift galaxies, the spectral synthesis technique in the rest-frame UV and Far-UV range will become one of a few number of tools remaining to study their young stellar populations in detail. The rest-frame UV lines and continuum of high redshift galaxies, observed with visible and infrared telescopes on Earth, can be used for accurate line profile fitting such as P\,{\sc{v}}$\lambda\lambda$1118,1128, C\,{\sc{iii}}$\lambda$1176, and C\,{\sc{iv}}$\lambda$1550. These lines are very precise diagnostic tools to estimate ages, metallicities, and masses of stellar populations. 

Here we discuss the potential for spectral synthesis of rest-frame UV spectra obtained at the Keck telescope. As an example, we study the 8 o'clock arc, a lensed galaxy at z=2.7322. We show that the poor spectral type coverage of the actual UV empirical spectral libraries limits the age and metallicity diagnostic. In order to improve our knowledge of high redshift galaxies using spectral synthesis, UV stellar libraries need to be extended to obtain accurate age, metallicity, and mass estimates likely to be occuring in young stellar populations observed in the early universe. 

\end{abstract}

\firstsection 
\section{Introduction}

The study of young stellar populations has benefited greatly from the ultraviolet (UV) evolutionary spectral synthesis technique, well-developed in the past two decades. In the 1000-2000\,\AA\ range, many lines display  P\,Cygni profiles typical of strong and dense stellar winds. The work by \cite{rob93} and \cite{lei95} have used IUE and HST/STIS UV spectra to show how the UV line profiles such as Si\,{\sc{iv}}$\lambda$1400, C\,{\sc{iv}}$\lambda$1550, and He\,{\sc{ii}}$\lambda$1640 were sensitive to the spectral type, luminosity class, and metallicity of early-type stars. The authors made used the line profiles observed in IUE and HST/STIS spectra to build an empirical stellar library for stellar population evolutionary models. This tool opened the door to the precise determination of the main physical properties of young stellar populations such as age, stellar mass, metallicity, and mass function (\cite[Leitherer \etal\ 1999]{lei99}). Later on, the UV range below Ly$\alpha$ became accessible with FUSE and more UV features were added to the collection of line profiles useful for evolutionary spectral synthesis (\cite[Pellerin \etal\ 2002]{pel02}, \cite[Robert \etal\ 2003]{rob03}). Good diagnostic lines found in this regime are C\,{\sc{iii}}$\lambda$1176, and P\,{\sc{v}}$\lambda\lambda$1118, 1128. 

So far the use of UV line profiles has proven to be quite successful to study young stellar populations in nearby galaxies (e.g. \cite[Pellerin 2006]{pel06}, \cite[Pellerin \& Robert 2007]{pel07}), and the future of UV spectral synthesis is expected to be quite bright as well. Indeed, the rest-frame UV spectra of high-redshift galaxies are now accessible with current optical facilities and many more of them will be observable with the coming generation of 30 meter class telescopes.

\section{10m Class Telescope Capabilities}

Although it is currently laborious and expensive to obtain UV rest-frame spectra of distant star-forming galaxies, a few works have already shown that is was possible to observe UV lines of young stellar populations with a spectral resolution good enough for stellar population synthesis studies. One impressive piece of work by \cite{sha06} presented the Keck optical spectra of fourteen z$\sim$3 galaxies. To gather the data, the authors integrated from 8 to 17 hours per galaxy in order to achieve a reasonable signal-to-noise ratio. Because of the high redshift, the authors identified several rest-frame UV lines, including the stellar lines of the Lyman series, Si\,{\sc{iv}}$\lambda$1400, and C\,{\sc{iv}}$\lambda$1550. Even if they were not clearly identified by the authors, stellar lines of C\,{\sc{iii}}$\lambda$1176, and even the fainter feature of P\,{\sc{v}}$\lambda\lambda$1118, 1128 in some cases, can be seen. Unfortunately, a more detailed and quantitative analysis of the stellar line profiles was not presented by the authors, likely because it was beyond the scope of their work. Nevertheless, their data clearly show the possibility of applying the UV spectral synthesis technique to high-redshift galaxies observed with 10m telescopes.

Another work by \cite{rix04} show an optical spectra of the z=2.7276 lensed galaxy MS\,1512$-$cB58 for which they performed spectral synthesis. An interesting aspect to the synthesis work was the use of a theoretical stellar library, allowing the use of lines free of interstellar medium (ISM) contribution and low metallicities. For this reason, the authors focused their work on lines less likely to be affected by ISM contamination, such as the 1425 index (a blend of Si, C, and Fe photospheric lines) and the Fe\,{\sc{iii}} blend around 1978\AA. The synthesis of more traditional lines wind sensitive lines of Si\,{\sc{iv}} and C\,{\sc{iv}} was therefore put aside due to the known contribution of ISM lines.

\subsection{The 8\,O'Clock Arc}
\label{sec8arc}

More recently, the rest-frame UV spectra of a lensed galaxy at z=2.7322 was obtained at the Keck telescope and was studied in detailed (\cite[Finkelstein \etal\ 2009]{fin09}). Because the galaxy was lensed, a modest integration time of 20 minutes was sufficient to detect absorption features. Stellar lines of the Lyman series, Si\,{\sc{iv}}$\lambda$1400, and C\,{\sc{iv}}$\lambda$1550 were identified by the authors and used to determine a precise redshift value. 

\begin{figure}[t]
\hspace*{-1.0 cm}
 \includegraphics[width=5.9in]{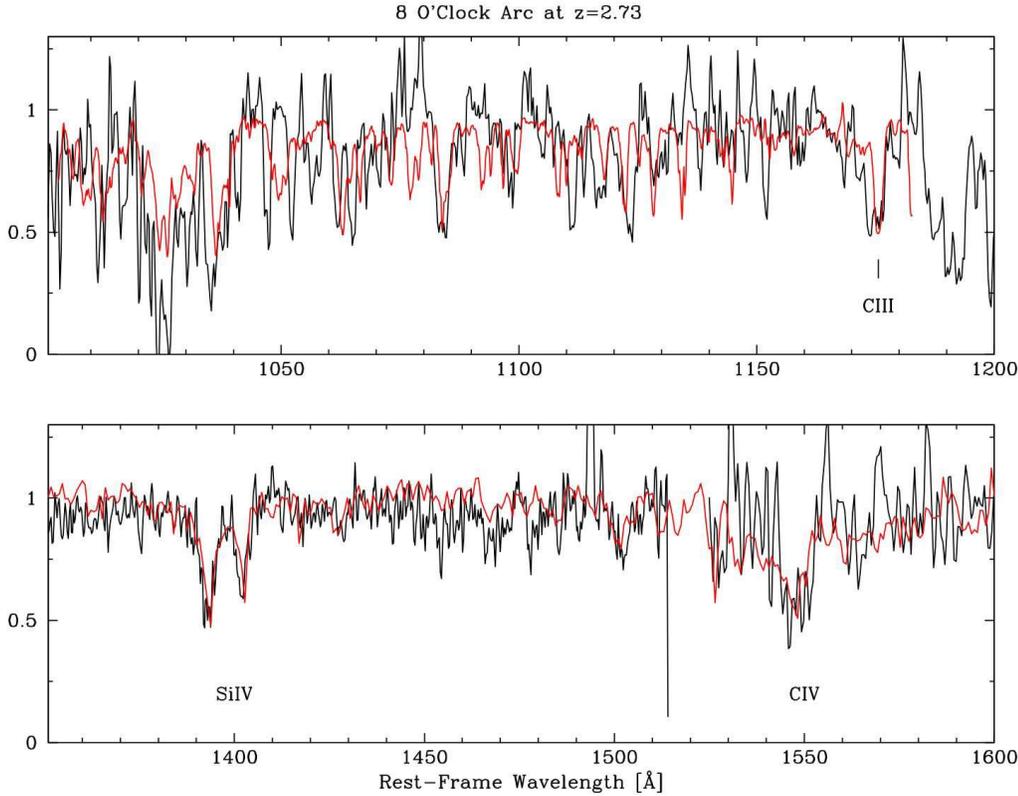}
 \caption{UV spectral synthesis of the lensed galaxy 8\,o'clock arc at z$=$2.7322. Black line: Keck/LRIS spectra. Red line: Best-fit model of an instantaneous burst of 7\,Myr at LMC metallicity using LavalSB (upper panel) and of a 10\,Myr burst at subsolar metallcity using Starburst99 (lower panel). The stellar lines of C\,{\sc{iii}}, Si\,{\sc{iv}}, and C\,{\sc{iv}} used for the line fitting are identified. }
   \label{arc}
\end{figure}

The Keck/LRIS spectrum of the 8\,o'clock arc galaxy is presented in Figure~\ref{arc} (black line). Details on the observation and reduction of the data can be found in the paper by \cite{fin09}. The spectrum includes the A1 and A3 comonents. In Figure~\ref{arc} we present a detailed analysis of the stellar line profiles. First of all, the detection of a thermal continuum below Ly$\alpha$ indicates that the galaxy hosts stars at least as hot as the B-type because A-type stars don't produce a significant flux in this range. The lines of C\,{\sc{iii}}, Si\,{\sc{iv}}, and C\,{\sc{iv}} are very sensitive to the age of the stellar population. For these lines, no wind profile was detected, meaning that there is no significant amount of O- and early B-type (i.e. B0-B2) stars in the A1 and A3 components. We fitted stellar population models on the stellar lines with Starburst99 (\cite[Leitherer \etal\ 1999]{lei99}) and LavalSB (\cite[Robert \etal\ 2003]{rob03}) to quantify the young stellar population properties. The 8\,o'clock arc spectrum agrees better with LMC metallicity models. Compared to the data, a solar metallicity model would have created a C\,{\sc{iii}} lines too deep and a SMC metallicity model would have lead to a line too shallow. Based on the work of \cite{pel07}, the LMC stellar library can fit relatively well metallicities from about 0.3 to 0.8\,Z$_{\odot}$, implying that the metallicity of the 8\,o'clock arc is likely within this range. This is consistent with the gas metallicity of 0.7$\pm$0.3\,Z$_{\odot}$ found for component A3 by \cite{fin09}. Unfortunately, the lack of stars cooler than B2-B3 in the far-UV empirical library at sub-solar metallicities does not allow to create a model for populations older than 7-10\,Myr without diluting the stellar lines. Therefore, we simply over-plotted the model of a 7\,Myr in the upper panel of Fig.~\ref{arc}. The best-fit age for the Si\,{\sc{iv}} and C\,{\sc{iv}} is around 10\,Myr. Here again, the lack of cooler B stars in the IUE/STIS library at sub-solar metallicities generates limitations on the age determination. A 10\,Myr burst is used in the lower panel of Fig.~\ref{arc}. The depth of the observed C\,{\sc{iv}} line is more consistent with sub-solar metallicity models. 

In conclusion, the UV spectral synthesis of the 8\,o'clock arc spectrum indicates the presence of a young stellar population of $\sim$10\,Myr with a metallicity between 0.3 and 0.8\,Myr. Due to the limitations in the empirical UV stellar libraries, this age may be slightly older. The absence of P\,Cygni profiles means that no hot O- or early B-type stars are presence. This is generally consistent with the work of \cite{fin09} based on SED synthesis. It also demonstrate the great potential of the UV spectral synthesis to quantify stellar population properties in star-forming galaxies at high-redshift.

\section{Ly$\alpha$ Emission Line as a Poor Age Diagnostic}

The strength of the Ly$\alpha$ emission line produced by a young stellar populations with hot OB stars can be estimated from the number of ionizing photons produced by those stars. It can be tempting to use the Ly$\alpha$ line to estimate the age of the population and is star formation rate because the feature is strong and easy to observe. However the interaction of Ly$\alpha$ photons with the surrounding ISM can lead to inaccurate Ly$\alpha$ line strengths. As first proposed by \cite{neu91} and recently revisited in detail by \cite{ver08} \cite{fin08}, the uniform distribution of gas and dust, or a multiphase ISM can either weaken or enhance the equivalent width of the Ly$\alpha$ emission line due to resonance scattering. It becomes obvious that one should be careful when using Ly$\alpha$ as an age diagnostic of a tool to estimate the star formation rate (see also e.g \cite[Gronwall \etal\ 2007]{gron07}). 

\section{The Future of UV Spectral Synthesis with the 30m Class Telescopes}

With the coming generation of 30 meter class telescopes such as the Giant Magellan Telescope (25m), the European Extremely Large Telescope (42m), and the Thirty Meter Telescope (30m), non-lensed high redshift galaxies (z$\ge$2) will be more accessible spectroscopic targets and a larger sample will be gathered for cosmological and galaxy evolution studies. The observed rest-frame spectra of those objects will be in the UV. Their photometric study over a wide wavelength range for SED fitting will become more and more difficult as the redshift increases and SED fitting will become more difficult and less reliable. The UV spectral synthesis will become a tool of choice to study the young stellar population properties (age, mass, metallicity, mass function) at high redshift.

However, the current empirical UV stellar libraries are based on stars from our own Galaxy for the solar metallicity library, and from the Magellanic Clouds for sub-solar metallicity libraries, limiting the metallicity range covered by the libraries. Furthermore, the lack of UV spectra of stars fainter than $\sim$B2-B5 in the sub-solar metallicity libraries will become a more serious issue because it leads to an upper limit to the age one can derived from synthesis models (see \S\ref{sec8arc}). 
To work around the metallicity issue, one needs to consider theoretical libraries (e.g. \cite[Rix \etal\ 2004]{rix04}). At this point, in order to get ready for the 30m class telescopes, we need to improve the sub-solar metallicity empirical libraries to extend the age range covered by the models. To extend the metallicity range coverage, we will need to keep improving theoretical libraries.

\end{document}